# VSC-WebGPU

A Selenium-based VS Code Extension For Local Edit And Cloud Compilation on WebGPU

Hao Bai
University of Illinois at Urbana Champaign
Champaign, Illinois, USA
Haob2@illinois.edu

*Abstract*—With the rapid development of information transmission, Software as a Service (SaaS) is developing at a rapid speed that everything originally local tends to be transplanted onto servers and executed on the cloud. *WebGPU* is such a SaaS system that it holds the GPU-equipped server to execute students' CUDA code and releases the *RESTful* front-end website for students to write their code on. However, programming on an HTML-based interface is not satisfactory due to a lack of syntax highlighting and automatic keyword complement. On the other side, *Visual Studio Code* is now becoming the most popular programming interface due to its strong community and eclectic functionalities. Thus, we propose such a system that, students write code locally using *VS Code* with its coding-auxiliary extensions, and push the code to *WebGPU* with only one button pressed using our *VSC-WebGPU* extension. The extension is divided into 4 parts: the login process for automatically logging the student into *WebGPU*, the pull process that pulls the code down to the local workspace, the push process that copies the code to the browser for compiling and running, and the exit process to exit the browser and close the connection. This 4-step architecture is also applicable for any other automated tools to push local code to authorization-required SaaS systems using Web automata.

## I. INTRODUCTION

Nowadays, the network is developing at a stunning speed, which makes "everything can be done through the network" less and less a dream. *Software as a Service*, or *SaaS*, is a product of such a bloom of information transmission. As its name has already expressed, *SaaS* is a delivery pattern that, anytime a user wants to utilize it, he can immediately do so, because all software just serves as services [1]. However, according to Cusumano [2], although *SaaS* and cloud computing replaces some traditional software products, they will not eliminate the software products anytime soon. One reason is that for a large amount of data, it still takes a considerable amount of time to upload and download. The other reason is that many of the local software have developed a complete community and ecology that is hard to challenge in a short time.

*WebGPU* is such an instance. Developed by the *IMPACT* group of UIUC, it is a *SaaS* system for students in the course *ECE-408* to write their CUDA C++ code, submit them, and get feedback [3]. After coding on the HTML page, students can click on the `Compile & Run` button so that the code will be submitted to the server, and the server will automatically run the code to find if there are any problems inside the code.

This is a very brilliant design for teaching because the university cannot guarantee that every student has a CUDA-based GPU on their laptop. With a high-end GPU running on the server, students can work on their code anywhere at any time - they can even work on their mobile phones. However, as coding on HTML pages provides little functionalities, especially for lack of syntax highlighting code tabbing (or referred to as automatic keyword complement), students have been complaining about the bad experience when using the web interface.

Thus, we introduce such a method that, users code locally, and use an automaton to push the code to the web interface, and then use the automata to get the standard output or error from the interface to the local interface.

## II. RELATED WORKS

### A. The Selenium Automaton

The *Selenium* automaton was created in 2004 by Jason Huggins as a tool to test Websites automatically, and it is now becoming one of the most popular testing tools in some famous companies like Google, Amazon, Microsoft, etc. It is also used for *Scrapy* programs to retrieve information from the Internet on servers [4].

Technically speaking, *Selenium* is an automaton that launches the browser driver and executes the pre-defined commands on the driver serially. All drivers have two modes: heading mode and headless mode. The headless mode means the browser is visible to users, and the headless mode means that the browser launches as a daemon process, which is not visible to users [5].[1]

*Selenium* has been released to different language platforms like *Python*, *JavaScript*, *Java,* and *PHP* [6]. It facilitates every language to simulate exactly what a human may do on browsers and is widely used for workers to submit work without operating on the website themselves.

Because *Selenium* has to deal with browsers, it has to deal with the potential delays and failures in HTTP requests. This leads to two choices: *blocking* running and *non-blocking* running. Blocking running means that the system must wait for the current command to finish the request (or `promise` in *node.js*) and then it goes on with the next command, while non-blocking running means that the system goes on to the next command even if the current request hasn't finished [7]. For language example, *Python* is a language that supports more blocking running while *node.js* supports both due to language characteristics.

---
[1] https://www.selenium.dev/selenium/docs/api/javascript/

In our extension, *Selenium* is based on *JavaScript* and executed on the *node.js* engine.

### B. The Visual Studio Code Extension

Recently, *Visual Studio Code* has become one of the most popular editors for programmers due to its high flexibility, high compatibility, strong community, and extension system. *VS Code* allows every programmer to develop their extensions and use others', achieving a situation that it becomes a great open-source community.

Specifically, an extension is a program in *VS Code* that is triggered whenever a specified condition is fulfilled.[2] There are many types of triggers, like pressing specified keys on the keyboard or typing in some keywords pre-defined in the extension. This is because *VS Code* is virtually a huge browser consisting of a large number of *HTML*, *CSS,* and *JavaScript* files, which makes itself a natural carrier of the event-driven editor.

There is a choice for *VS Code* extensions: developers can use *JavaScript* or *TypeScript* for preference. *TypeScript* provides great compatibility for code editing functionalities and almost all functionalities in *JavaScript*, while *JavaScript* provides lower and faster support for system IO, network manipulation, and also asynchronization patterns

In our extension, we used *JavaScript* because it's enough to implement our requirements.

### C. The Node.js Engine

The *VS Code* extension uses *node.js* as its engine. In other words, it uses *node.js* as its compiler for the source code of the extension. *Node.js* is a cross-platform, open-source running environment on the server-side and it's a part of the *Linux Foundation*.

As a running environment mostly for *JavaScript*, *node*.js provides a great performance on event-driven and asynchronization IO manipulation [8].[3] As mentioned in Section 2.2, *VS Code* extensions are virtually small programs with triggers as events, *node.js* provides a natural advantage to run for the extensions.

*Node.js* becomes more of a great choice when considering *Selenium*. As mentioned in Section 2.1, *Selenium* is an automaton dealing with Internet problems so it has to consider network connection issues, which leads to the question of "blocking or not" [9].

Overall speaking, a non-blocking system is much slower than a blocking system, because it has to wait for a `promise` to be sent back to carry on the next command. Non-blocking systems, on the other side, provide an asynchronization choice for programmers, that they can use keywords like `async` to announce that the run-time head does not need to wait for the promise to be sent back, and when the promise is sent back, the head calls a *callback* function [10]. This idea implies a multi-threading possibility and improves the performance of the system to a large extent.

Hitherto, we've finished all the prerequisites for this extension: it's a *Selenium*-based *VS Code* extension running on the *node.js* environment.

### III. EXTENSION ARCHITECTURE

The basic architecture of this *VS Code* extension is flat with 4 operations: `login`, `pull`, `push`, and `exit`. The combination of functions of the 4 operations forms a bijection relationship with the set of all the functionalities we'll use in the process of using WebGPU for coding.

Firstly, after the extension is activated, it will automatically run code in the `__start()` function, which includes all the initializations for the extension. All later functions are event-driven, so users can use a keypress or mouse click to call those utility functions. The graphical illustration is shown in Figure 1.

The difference between the *functional subroutine* and the *structural subroutine* is that *functional subroutine* is executed only once in a function and contains utilities, while *structural subroutine* is often executed multiple times in a function and always contains structural requisites.

This flat and separate workflow for the extension provides flexible usage for users. The following part contains the detailed things we do in each function. Because the `exit()` function is simple, we'll not mention it below.

### A. Initialization

Before all the operations, work needs to be done to initialize the expansion for work. After the user activates the `login()` function, the `__start()` process will be called automatically before the system virtually executes the `login()` operation.

In the `__start()` process, the extension loads the configuration into the system and localizes them as local variables. The process will also create an output channel called `WebGPU` on the *VS Code* interface for later possible outputs. At last, the system starts the appointed browser driver (such as Chrome, Firefox, and Microsoft Edge) and waits for utility functions on the driver.

### B. The Login Operation

The `login` operation is the process for simulating users to log in to the WebGPU website and access the required lab number. The overall logic is pretty simple, but there are some potential vulnerabilities inside this process driver.

---

[2] https://code.visualstudio.com/api
[3] https://nodejs.org/en/

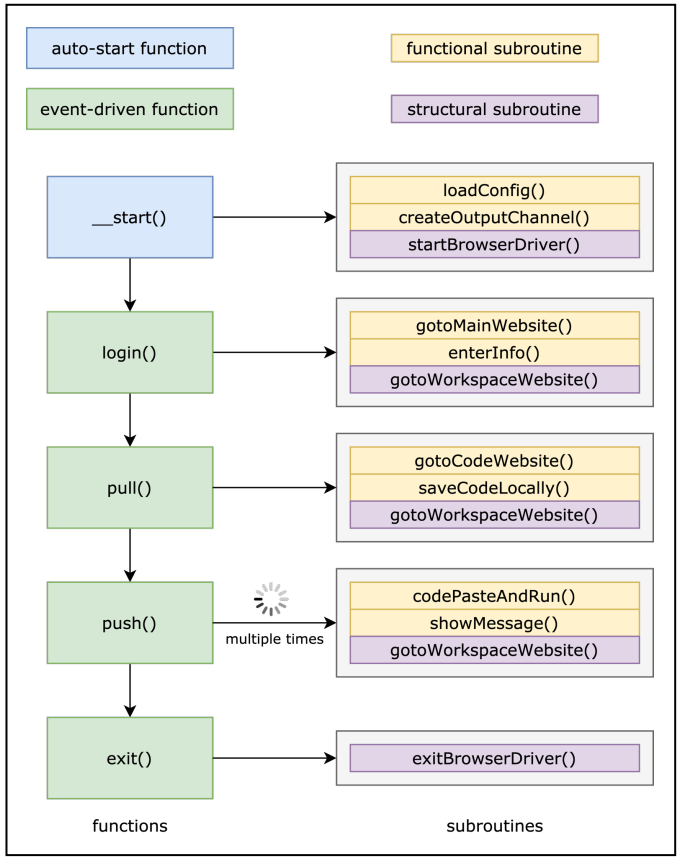

Figure 1. The overall structure of the extension. **Left**: the functions implemented in the extension. **Right**: the subroutines (functionalities) of each function. Note that the `push` function may be called several times, because students tend to write code and run several times.

When logging in, both account and password need to be delivered to the website. In this case, users should not be asked to type in their information each time they log in, so we took the approach to store their information in the config file and local it whenever needed. Because this information will hardly change in one semester, this approach is much more efficient.

The `gotoMainWebsite()` subroutine also contains potential challenges that there may be *SSL failures* on the website server from time to time. In this case, we add an auto-detect system that, if there is such a problem, we'll instruct the browser driver to ignore it and go on, which is equivalent to two mouse-click operations.

### C. The Pull Operation

The `pull` operation pulls the latest version of code on the server down to the currently opened window in *VS Code*. This pattern is applied because sometimes programmers like to deal with one bunch of code at a time. Thus, if they forget to save their file on the local device, they can still get their last storage on the server when they start up next time.

Technically speaking, the `pull` operation utilizes the *WebAPI* `get()` to obtain the code from the *RESTful API* (REpresentational State Transfer Application Program Interface) [11] of the *WebGPU* backend. This approach is available because the back-end and front-end of *WebGPU* are also separate, as shown in Figure 2.

### D. The Push Operation

The `push` operation is the most complicated one in the extension. It copies the code in the *VS Code* workspace and paste it onto the browser driver, and click the Run button on the web interface. Then it takes the message (whether it's an error or an output) back to the *VS Code* output channel WebGPU. Respectively, it contains two functional subroutines: `codePasteAndRun()` and `showMessage()`.

In the `codePasteAndRun()` subroutine, a technical challenge is that there is no *RESTful API* (refer to Figure 2), so we need to copy and paste the code to the website interface through the browser driver directly. However, the design pattern of the website interface does not allow pasting. We will discuss this problem as an individual thread in Section 4.3.

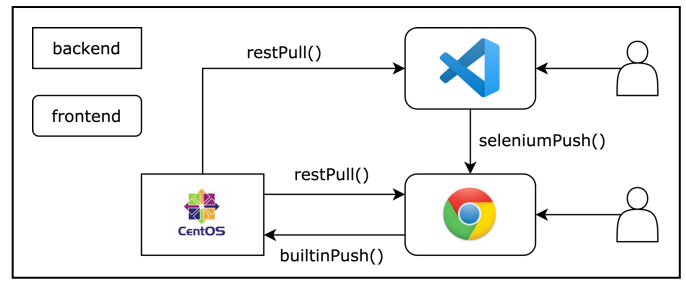

Figure 2. The `pull()` and `push()` process of the *WebGPU* system. The system is divided into front-end and back-end. Because the original `pull()` process utilizes the *RESTful API*, we can directly use it to get the code. However, because the original `push()` process does not provide any APIs, we can only copy the code from the local workspace to the browser driver first, as will be discussed in Section 3.4.

In the `showMessage()` subroutine, we need to only download useful information. For example, a copy of the code being running is not necessary, but it appears on the website interface. In this case, we need to specify the `xpath` of each element shown so that we can filter information.

## IV. TECHNICAL DETAILS

In this part, we discuss the technical details in our extension, including the bugs we met and the issues we resolved. Although the code is short enough (about 300 lines), it includes a large amount of knowledge including choice of blocking and non-blocking using the `promise` object in each case, implicit waiting and explicit waiting, the "train method" for an activation event. Our code is available on GitHub[4]. If you have any issues with the extension, please leave a message on the *issue* part of this repository.

### A. Choice of Blocking and Non-blocking Running

The choice of a blocking or non-blocking design is pretty common in this extension. The principle is: if the next command must be executed after the current command has the `promise` returned, then using a non-blocking running mode is unacceptable. Please refer to Figure 3 for example. The function `function()` calls a function with `request()` command with a `Promise` and wants to give feedback to the console. When this is the case, programmers need to choose **A** as the feedback announcement, because choice **A** only prints the feedback after the `request` function is finished and ready to return the Promise. If we mistakenly use **B** as the choice, no matter what the `callback_function()` returns (`Promise` successfully returns or not), the `print()` command will always be executed immediately after the `callback_function()` is called, and thus lose its proper functionality.

---
[4] https://github.com/BiEchi/ece408-remote-control

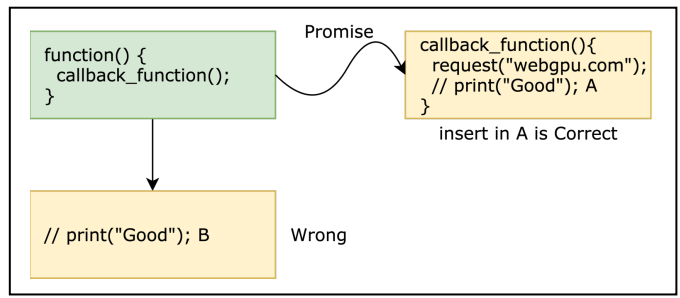

Figure 3.  A comparison of blocking or not. Server programmers tend to use choice **A** to report potential errors. **A**: blocking. **B**: non-blocking.

The blocking concept is used not only in web programming. In Operating System programming and any tasks with asynchronization or callback functions, this concept is rather important for correctly manipulating one function and another.

### B.  Implicit Waiting and Explicit Waiting

The concepts of implicit waiting and explicit waiting are also pretty common in `http` requests. Overall speaking, implicit waiting waits for all the web pages to be loaded to locate the specific HTML element, but explicit waiting waits for exactly the specified HTML element to be loaded, and will immediately return the element as soon as the element has been loaded to the website DOM.

To memorize the two terms, think of such a case - your friend is waiting for you to attend a party. In one case, he is waiting for a bunch of people including you, but he cares about you most, so actually, he is "implicitly" waiting for you as he not only waits for you. In the other case, he is saying that "I'm only waiting for you" so he is actually waiting for you and only you, so it would be very personal and called "explicit" waiting.

In *Selenium*, the discussion of the pros and cons of the two waiting methods has been pretty popular, but the rules for the two methods are virtually the same for blocking and non-blocking running. Please refer to Figure 4 for example. To load the `target` element, the implicit waiting method waits for all the elements to be loaded into DOM, while the explicit waiting method only waits for the necessary elements to be loaded, i.e., all the elements that this element is cascaded on.

Technically speaking, explicit waiting is much faster, but implicit waiting is much safer. It's pretty easy to understand why explicit waiting methods are faster - because it does not wait for the non-relevant elements to be loaded. However, in some cases, it's hard to define what is "non-relevant", because some websites require some particular elements to be loaded as a trigger for later operations. Thus, the choice of the two methods is not absolute.

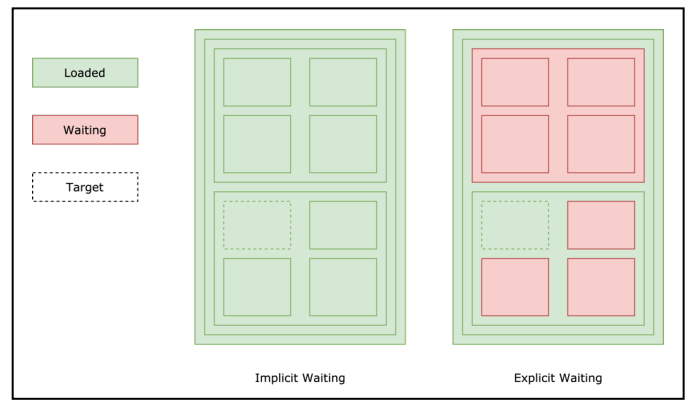

Figure 4.  Difference between implicit waiting and explicit waiting. The explicit waiting method waits for only the target to be loaded to carry on the next command (non-blocking) while the implicit waiting method waits for all the elements in the web page to get loaded (blocking).

### C.  The "Train Method" for Combinational Inputs

Basically, all the commands that we do on *Selenium* need to have a subject. Different objects has different commands to control, and calling a command that does not exist in an object is not acceptable. Please refer to Figure 5 for example. We can click a button while we cannot send keys to a button. Meanwhile, we can send keys to an input box while we cannot click the input box.

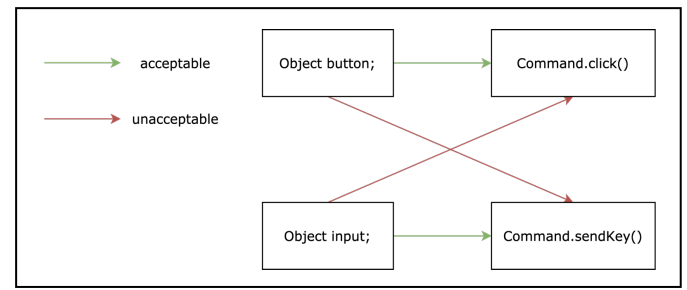

Figure 5.  We cannot send a key to a button, and we cannot click the input box.

In this case, the issues come out of the WebGPU code editing site. Our goal is to send our code to the code editing page using the `sendKey()` approach by entering `ctrl+v`. However, this site uses a *RESTful* approach to render colors on code, and thus it is a common `<a>` element instead of an `<input>` element, so we cannot send keys to the element.

```
const actions = driver.actions();
actions
.click(code_line)
.keyDown(Key.COMMAND)
.sendKeys('a')
.sendKeys('v')
.sendKeys('s')
.keyUp(Key.COMMAND)
.click(compile_button)
.click(all_button)
.perform()
.then(function(){
    feedback();
})
```

The solution is, we use a combinational input method called the "train" method, to send the keys directly to the event (or called the browser), instead of a specific element. The source code of this part is shown above.

In the code, we send all the commands into one event called `driver.actions()`, so all the commands including `keyDown()`, `sendKeys()`, `click()` can be executed serially like a "train" and none of these commands can be moved out of the "train". Compared to executing commands on an element, the "train" method provides a way to execute all the commands directly to the browser just like what is manually done.

After the "train" method is applied, we use a blocking function `feedback()` to deal with the events after all the train commands are executed. In this function, we also have something to do with detecting whether the compilation process is done so it's more about blocking running processes.

### D. The Heading Mode and Headless Mode

The difference between a heading mode and a headless mode is that the heading mode uses a browser that is visible to the users, while the headless mode makes the browser a pure daemon process and thus not visible to users anymore. By default, the driver is in the heading mode. To turn on the headless mode, the logic is to change the option when launching the browser driver before launching. The concrete code is shown below.

```
var options = new chrome.Options();
options.addArguments('--headless');
var driver = new webdriver.Builder()
    .forBrowser('chrome')
    .withCapabilities(options)
    .build();
```

Originally, our goal was to design such a headless mode so that all users do not need to care anything about how the browser works, but because we used the "train" method, we must use the heading mode. Because the "train" method needs to send keys directly to the browser using the keys on the machine, the browser must appear as a visible entity to be performed on. Yet we're still trying to achieve the headless mode because it's very friendly to users, and hopefully, we can manage it in another way.

## V. SCENARIO GENERALIZATION

In this part, we generalize our design to the larger world. As we've already discussed in Section 1, more and more delivery methods are turning to the kind that has a consistent website front-end and a running server backend, in which case our work can be applied.

The ideology here is to apply a browser or server automata and a local workspace with *VS Code*. The development process would be always the same: initialize, login, pull, push, and exit. The `login` process sets up the connection between the VS Code extension and the browser (or server), the `pull` process downloads the code to *VS Code* for local editing, the push process uploads the code file back to the browser or server for cloud computing, and show the returned message locally, and the `exit` process closes the connection. The graphical illustration is shown in Figure 6.

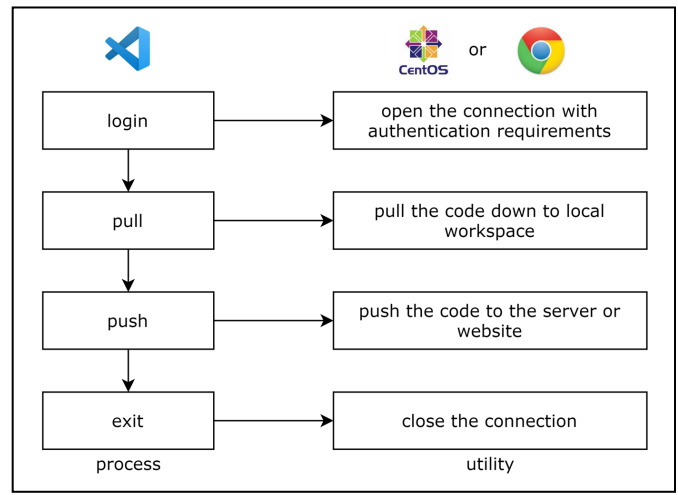

Figure 6. Our design pattern can use *VS Code* to create a connection between not only the browser but also directly the server. The workflow applies to most of the scenarios where we want to run our code on the cloud instead of a local machine.

## VI. CONCLUSION

In this paper, we introduced the project for users to code locally and use an automaton to push the code to the web interface, and then use the automata to get the standard output or error from the interface to the local interface. We discussed the project goal, the achievements we've already received, the issues we met when developing, and the relevant principles behind the code. Yet more work needs to be done on this extension to improve the robustness when a network connection fails.

ACKNOWLEDGMENT

I need to express my greatest thanks to Professor Kindratenko, who encouraged me to go on with this VS Code extension. I also need to thank ECE408 Teaching Assistant Andrew Schuh (also the maintainer of WebGPU), who provided me with the web API for the `pull` part of the extension. Gratitudes are also given to my peer student Zhenzuo Si, who encouraged me to develop such a tool.

## REFERENCES

[1] Godse, M., & Mulik, S. (2009, September). An approach for selecting software-as-a-service (SaaS) product. In 2009 IEEE International Conference on Cloud Computing (pp. 155-158). IEEE.

[2] Cusumano, M. (2010). Cloud computing and SaaS as new computing platforms. Communications of the ACM, 53(4), 27-29.

[3] Dakkak, A., Pearson, C., & Hwu, W. M. (2016, May). Webgpu: A scalable online development platform for gpu programming courses. In 2016 IEEE International Parallel and Distributed Processing Symposium Workshops (IPDPSW) (pp. 942-949). IEEE.

[4] Grigors, E. J. (2021). Beautiful Soup, Selenium and Scrapy Speed, Efficiency and Solution Comparison.

[5] XIAO, B., & CHEN, Z. X. (2008). Headless Mode Application Study Based on JAVA. Computer Knowledge and Technology, 2008, 26.

[6] García, B., Munoz-Organero, M., Alario-Hoyos, C., & Kloos, C. D. (2021). Automated driver management for selenium WebDriver. Empirical Software Engineering, 26(5), 1-51.

[7] Shah, H. (2017). Node. js challenges in implementation. Global Journal of Computer Science and Technology.

[8] Satheesh, M., D'mello, B. J., & Krol, J. (2015). Web development with MongoDB and NodeJs. Packt Publishing Ltd.

[9] Li, R., & Hu, J. (2013). Study of Asynchronous Non-Blocking Server Based on Nodejs.

[10] Li, R., & Hu, J. (2013). Study of Asynchronous Non-Blocking Server Based on Nodejs.

[11] Wang, S., Keivanloo, I., & Zou, Y. (2014, November). How do developers react to *RESTful* api evolution?. In International Conference on Service-Oriented Computing (pp. 245-259). Springer, Berlin, Heidelberg.